# Highly fabrication tolerant InP based polarization beam splitter based on p-i-n structure

NICOLÁS ABADÍA,[1,3,*] XIANGYANG DAI,[2] QIAOYIN LU,[2,4,*] WEI-HUA GUO,[2] DAVID PATEL,[3] DAVID V. PLANT,[3] AND JOHN F. DONEGAN[1]

[1]*Semiconductor Photonics Group, School of Physics and CRANN, Trinity College, Dublin 2, Ireland*
[2]*Wuhan National Laboratory for Optoelectronics, Huazhong University of Science and Technology, Wuhan 430074, China*
[3]*Department of Electrical and Computer Engineering, McGill University, Montreal, QC, H3A 2A7, Canada*
[4]*luqy@hust.edu.cn*
[*]*abadian@tcd.ie*

**Abstract:** In this work, a novel highly fabrication tolerant polarization beam splitter (PBS) is presented on an InP platform. To achieve the splitting, we combine the Pockels effect and the plasma dispersion effect in a symmetric 1x2 Mach-Zehnder interferometer (MZI). One p-i-n phase shifter of the MZI is driven in forward bias to exploit the plasma dispersion effect and modify the phase of both the TE and TM mode. The other arm of the MZI is driven in reverse bias to exploit the Pockels effect which affects only the TE mode. By adjusting the voltages of the two phase shifters, a different interference condition can be set for the TE and the TM modes thereby splitting them at the output of the MZI. By adjusting the voltages, the very tight fabrication tolerances known for fully passive PBS are eased. The experimental results show that an extinction ratio better than 15 dB and an on-chip loss of 3.5 dB over the full C-band (1530-1565nm) are achieved.

**OCIS codes:** (250.5300) Photonic integrated circuits; (130.3120) Integrated optics devices; (230.1360) Beam splitters; (230.5440) Polarization-selective devices.

## References and links

## 1. Introduction

Due to the increasing internet traffic, there is a need for upgrading current optical networks. One of the research lines to meet this demand is by implementing phase modulation formats that increases the number of bits per symbol in the optical link. An example is the Dual-Polarization Quadrature-Phase Shift Keying (DP-QPSK) combined with a digital coherent detection scheme [1]. For the coherent detection, both free space and highly integrated schemes have been proposed, with InP as a key integration platform [2].

A vital component of the coherent scheme is the polarization beam splitter (PBS). This device separates the TE mode from the TM mode within the integrated circuit. Several polarization beam splitters in both silicon [3–7] and indium phosphide (InP) have been demonstrated [8–11]. One of the major fabrication issues for the PBS in both Si and InP is that they have stringent fabrication tolerances in order to produce a high extinction ratio.

There are some reports of PBS in InP that try to relax or ease fabrication tolerances using different techniques such as Si/SiO$_2$ Periodic Layer Structures (PLS) [10] or thermal tuning [9,11,12]. In the case of the PLS, this adds complexity to the fabrication, requiring the deposition of many layers in a periodic stack. On the other hand, thermal tuning can be used to overcome fabrication tolerances by changing the current to the heater but the effect is weak. Furthermore, the thermal effect affects both the TE and the TM mode at the same time, and therefore, this PBS design is difficult to adjust [13].

To simplify the adjustment of the PBS a new method has been proposed in [13], this method exploits the Pockels effect in a multiple quantum well InP based structure. To produce the polarization splitting function they exploit the fact that the Pockels effect affects only the TE mode. Positive and negative changes to the refractive index can be engineered by changing the orientation of the device on the wafer. By reverse biasing the phase shifters that make up the PBS, the splitting function can be realized. This PBS can be easily adjusted by adjusting the bias on the phase shifters. Nevertheless, this device needs two phase shifters perpendicular to each other giving a relatively large layout of 1.5 mm x 2.5 mm per device.

As an alternative way to realize a PBS, we propose to combine the use of the plasma dispersion effect and the Pockels effect in a p-i-n structure with a bulk intrinsic region. Furthermore, with this approach, we do not need to place two phase shifters perpendicular to one another and therefore need less space on the wafer. The experimental results show a polarization extinction ratio better than 15 dB over the entire C-band can be obtained with an on-chip loss of around 3.5 dB. Such a device provides an easy adjustment in just two steps which consist of setting two different voltage to produce the splitting function. Also, by adjusting the voltages in forward and reverse bias, fabrication errors can be overcome.

## 2. Principle of operation

In this section, to explain how our PBS design works, we first summarize the exploited electro-optic effects on the InP-InGaAsP materials platform. Then the PBS based on a symmetric MZI is shown. Finally, we describe how to easily adjust the PBS to produce the TE/TM splitting function and to ease the fabrication tolerances.

### 2.1 Exploited Effects in InP-InGaAsP

The plasma dispersion effect changes the refractive index of InP-InGaAsP when carriers are injected into the structure [14,15]. It is exploited here in a p-i-n structure in forward bias. This effect is polarization independent, so, it affects both the TE and TM mode of an integrated waveguide. The change of the effective index of the modes supported by the integrated waveguide are denoted by $\Delta n_{eff, TE}(V_{forward}) \approx \Delta n_{eff, TM}(V_{forward})$, where $\Delta n_{eff, TE}$ ($\Delta n_{eff, TM}$) is the effective index change of the TE mode (TM mode) and $V_{forward}$ is the forward bias voltage applied to the p-i-n structure.

The Pockels effect is the linear change of the refractive index of the material when an electric field is applied to it [15]. It is present in InP-InGaAsP in a reverse biased p-i-n structure. This effect is polarization dependent affecting only the TE mode. The effective index change of the TE mode is $\Delta n_{eff, TE}(V_{reverse})$ where $V_{reverse}$ is the reverse bias voltage in the p-i-n structure. In an InP-InGaAsP p-i-n structure there may also be present the Kerr effect and the band-filling effect when it is in reverse bias [14]. In the structure considered in this paper, the Pockels effect is dominant in reverse bias since we use a bulk intrinsic layer in the p-i-n as opposed to the quantum well structure in [13] and the bandgap of the material is set at 1.072 eV (1156.6 nm) which is far from the 0.8 eV (1550 nm) working energy.

The wafer structure used in this work is presented in Table 1. There, the layer number, the material, the thickness of each layer, the dopant, the type of dopant and the carrier concentration is summarized.

Table 1. Structure of the wafer.

| Layer | Material | Thickness [nm] | Dopant | Type | Carrier [cm$^{-3}$] |
|---|---|---|---|---|---|
| 10 | InP | 20 | Silicon | N | $1 \times 10^{18}$ |
| 9 | Ga$_{0.11}$In$_{0.89}$As$_{0.24}$P$_{0.76}$ | 50 | Silicon | N | $1 \times 10^{18}$ |
| 8 | InP | 400 | Silicon | N | $1 \times 10^{18}$ |
| 7 | InP | 300 | Silicon | N | $5 \times 10^{17}$ |
| 6 | InP | 300 | Silicon | N | $3 \times 10^{17}$ |
| **5** | **Ga$_{0.2}$In$_{0.8}$As$_{0.42}$P$_{0.58}$** | **400** | **Undoped** | **-** | **-** |
| 4 | InP | 300 | Zinc | P | $3 \times 10^{17}$ |
| 3 | InP | 1000 | Zinc | P | $5 \times 10^{17}$ |
| 2 | Ga$_{0.47}$In$_{0.53}$As | 80 | Zinc | P | $1 \times 10^{19}$ |
| 1 | InP | 500 | Zinc | P | $1 \times 10^{18}$ |

We will use the plasma dispersion and Pockels effects in a symmetric Mach-Zehnder interferometer (MZI) to demonstrate the polarization beam splitter function. Layer 5 is highlighted, it is the intrinsic region in the p-i-n structure.

### 2.2 Mach-Zehnder interferometer based polarization beam splitters

The proposed PBS is based on a 1x2 symmetric MZI as represented in Fig. 1.

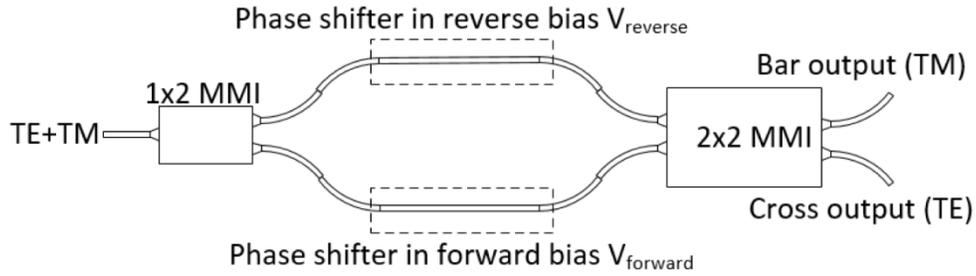

Fig. 1. Structure of the proposed PBS which consists of a 1x2 symmetric MZI.

Figure 1 illustrates a 1x2 symmetric MZI with one input and two outputs: one for the TE mode (bar output) and one for the TM mode (cross output). Each arm of the MZI consists of a phase shifter with a vertical p-i-n structure based on the wafer of Table 1.

To produce the polarization splitting function one arm of the MZI is driven in forward bias $V_{forward}$ (bottom arm in Fig. 1) while the other is driven in reverse bias $V_{reverse}$ (upper arm in Fig. 1.). The forward bias $V_{forward}$ will affect a change in the effective index of the TE mode $\Delta n_{eff,\,TE}(V_{forward})$ and the TM mode $\Delta n_{eff,\,TM}(V_{forward})$. On the other hand, the one in reverse bias $V_{reverse}$ will only affect the effective index change of the TE mode $\Delta n_{eff,\,TE}(V_{reverse})$.

The splitting function is done by having different interference conditions for the TE and the TM modes at the outputs of the MZI. One output has constructive interference for the TE mode and the other output has constructive interference for the TM mode as shown in Fig. 1.

To clarify the interference conditions, we write the equations of the phase difference seen by the TM and TE mode:

$$\Delta\varphi_{TM} = \frac{2\pi}{\lambda} \cdot \left(\Delta n_{eff,TM}\left(V_{forward}\right)\right) \cdot L + \varphi_1 \qquad (1)$$

$$\Delta\varphi_{TE} = \frac{2\pi}{\lambda} \cdot \left(\Delta n_{eff,TE}\left(V_{forward}\right) + \Delta n_{eff,TE}\left(V_{reverse}\right)\right) \cdot L + \varphi_2 \qquad (2)$$

where $\Delta\varphi_{TM}$ is the phase change seen by the TM mode and $\Delta\varphi_{TE}$ is the phase change induced in the TE mode; L is the length of the phase shifters in Fig. 1, and $\lambda$ is the free space wavelength. The terms $\varphi_1$ and $\varphi_2$ are fixed contributions that comes from fabrication errors, the MMIs, junctions, variation in the thickness of the layers, etc. Since the TM mode is only affected by the plasma dispersion effect, only $\Delta n_{eff,\,TM}(V_{forward})$ appears in Eq. (1). On the other hand, the TE mode is affected by both the plasma dispersion effect, $\Delta n_{eff,\,TE}(V_{forward})$ and the Pockels effect $\Delta n_{eff,\,TE}(V_{reverse})$ as given in Eq. (2). By changing $\Delta n_{eff,\,TE}(V_{reverse})$ the interference condition for the TE mode can be different from the interference condition of the TM mode.

To adjust the PBS to produce the splitting function it is necessary to experimentally find the forward bias voltage $V_{forward}$ in one arm of the MZI and the reverse bias voltage $V_{reverse}$ in the other arm of the MZI as shown in Fig. 1.

## 2.3 Adjustment of the PBS

To adjust the PBS, first we inject a TM mode and forward bias one arm. There will be one voltage called $V_{forward}$ in which the TM mode will be output completely to the cross output of the 1x2 MZI in Fig. 1. We fix this arm to $V_{forward}$.

Then, a TE mode is injected while we reverse bias the other arm. There will be a voltage $V_{reverse}$ in which the TE power will exit the cross output of the 1x2 MZI in Fig. 1 hence, producing the splitting function.

One advantage of the proposed PBS is that it can be fabrication tolerant since the phase changes induced by fabrication errors (e.g. variation in the thickness of the layers through the wafer, variations in the width of the waveguides, etc.) in the MZI can be compensated by changing both $V_{forward}$ and $V_{reverse}$ while adjusting the device.

## 3. Experimental results

The fabricated structure is shown in Fig. 2. It consists of an input waveguide, a 1x2 MMI (10 μm x 106.5 μm), the two MZI arms (2.5 mm), and a 2x2 MMI (15 μm x 317.8 μm). The widths of the waveguides used is 2 μm. The outputs of the 1x2 MZI are tapered to facilitate alignment when using the end-fire technique. On top, there is the DC ground gold contact. The other two contacts are the contacts of the two arms of the MZI. The contacts of the arms were segmented along the vertical p-i-n phase shifters (over the waveguide) to reduce the insertion loss of the device which affects the TM mode in particular. At the bottom of Fig. 2, a straight waveguide is placed to facilitate the measurement of the insertion loss of the device and normalize the transmission measurements.

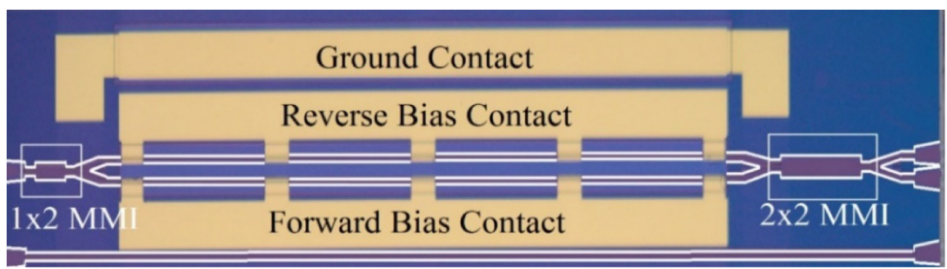

Fig. 2. Fabricated structure of the PBS with a straight waveguide at the bottom for normalization of the insertion loss.

The setup used to characterize the device is represented in Fig. 3. It consists of a tunable laser (TL) and a bare fiber rotator (FR) to set the input polarization of the light. Polarization maintaining lensed fibers (PMF) are used to excite the circuit. We align the slow axis of the polarization maintaining waveguide parallel and perpendicular to the integrated waveguide to inject a TE and a TM mode. The output of the PBS is measured using a power meter (PM). To set $V_{forward}$ and $V_{reverse}$ two power supplies (PS1 and PS2) were used.

To align the slow axis of the polarization maintaining fiber with the input waveguide of the PBS we maximized the extinction ratio in reverse bias by scanning $V_{reverse}$ and rotating the fiber rotator. When the extinction ratio in reverse bias reaches maximum we assume that the slow axis of the input waveguide is parallel to the surface of the chip and that we are injecting a TE mode. The maximum extinction ratio is reached once a TE mode is injected since the Pockels effect only affects the TE mode. This gives us a certain position in the fiber rotator. To inject a TM mode, we rotate the fiber rotator 90 degrees from that position. Once we know the two positions of the fiber rotator we proceed to the measurement of the PBS. This setup is not a common one for waveguide device measurements. It was made necessary since the devices were placed at the center of the carrier rather than at its edge. Consequently, it was not possible to excite it with free space optics.

Since the plasma dispersion effect affects both the TE am TM mode and the Pockels effect affects only the TE mode, we first inject a TM mode to determine $V_{forward}$. The result is shown in Fig. 4.

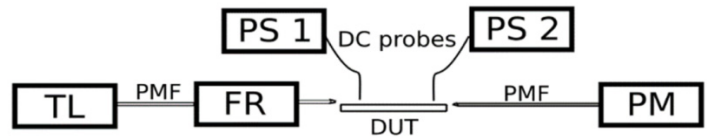

Fig. 3. Setup to characterize the PBS. It consists of two power supplies (PS1 and PS 2), one tunable laser (TL), one bare fiber rotator (FR), and a power meter (PM). Polarization maintaining lensed fibers were used at the input and at the output of the Device Under Test (DUT)

It is possible to observe that the TM mode is going to the bar output for a voltage around $V_{forward}$ = 1.325-1.45 V. Due to a small wavelength dependence of the plasma dispersion effect the voltage $V_{forward}$ is slightly different depending on the wavelength. The polarization extinction ratio of the PBS is in the range 15-20 dB depending on the wavelength and the on-chip loss is around 3.5 dB.

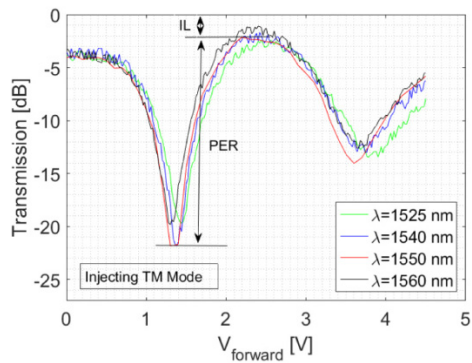

Fig. 4. Bar output of the PBS when injecting a TM mode at the input. One arm is driven in forward bias $V_{forward}$ from 0 to 5 V. Several wavelengths are measured in the C-band to demonstrate the bandwidth of the device. PER stands for polarization extinction ratio and IL for insertion loss.

Once $V_{forward}$ is known, we inject a TE mode and we reverse bias the other arm at $V_{reverse}$. The results are shown in Fig. 5. In this case, the polarization extinction ratio is around 15-18 dB depending on the wavelength and the on-chip loss is around 3.5 dB.

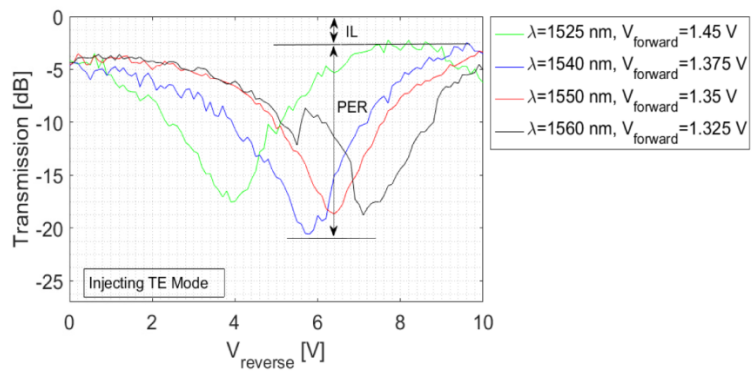

Fig. 5. Cross output of the PBS when injecting a TE mode at the input. While one arm is driven in forward bias $V_{forward}$ the other arm is reverse biased $V_{reverse}$ from 0 to 10 V. Several wavelengths are measured in the C-band to demonstrate the bandwidth of the device. PER stands for polarization extinction ratio and IL for insertion loss.

In the case of Fig. 5, it is possible to observe that the maximum polarization extinction ratio depends on the voltage $V_{reverse}$. The minimum happens between $V_{reverse}$ = 4.5-6 V depending on the wavelength. This seems to be in contradiction with the fact that the Pockels effect is wavelength insensitive when the wavelength is far from the bandgap of the material. However, due to a fabrication issue with this first round of devices, there is not good electrical isolation between the arms of the MZI. We believe that this issue can be solved in future fabrication runs.

Finally, we present the polarization extinction ratio of the PBS versus wavelength. It is shown in Fig. 6 is based on the measurements of Fig. 4 and 5. It is possible to observe that the PBS can work across all the C-band with a polarization extinction ratio better than 15 dB.

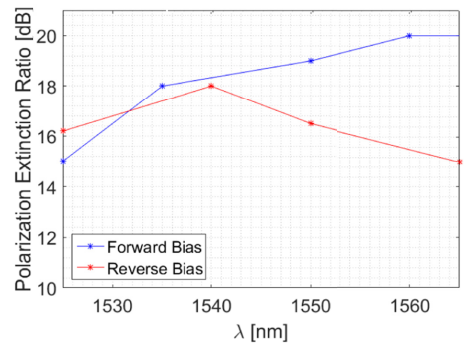

Fig. 6. Polarization extinction ratio of the PBS in forward and reverse bias versus wavelength

## 4. Comparison with the state-of-the-art

In this section, the proposed PBS presented in this work is compared with the current state-of-the-art in Table 2.

Table 2. Summary of the state-of-the-art of PBS based on InP platform where PER stands for Polarization Extinction Ratio, IL for Insertion Loss, BW for Bandwidth, and L is the length of the device.

|  | TU/e [8] (2007) | TU/e [16] (2007) | Malaga [10] (2013) | Fraunhofer [11] (2014) | Mitsubishi [17] (2015) | NTT [13] (2016) | This work (2017) |
|---|---|---|---|---|---|---|---|
| PER | 10 dB | 13 dB | 13 dB | 25 dB | 15 dB | 14 dB | 15 dB |
| IL | 5 dB | 3.6 dB | - | 2.5 dB | 2-4 dB | - | 3.5 dB |
| BW | - | 45 nm | 30 nm | - | 21 nm | - | 35 nm |
| L | - | 1325 µm | 1.5 mm | 2 mm | 370 µm | 1.5 mm x 2.5 mm | 0.5 mm x 2.5 mm |

The PBS proposed in this work has a polarization extinction ratio better than 15 dB across the whole C-band, offering a better performance than [8,9,12,16], even though [16] offers similar insertion losses and a better bandwidth of 45 nm it cannot be calibrated to avoid fabrication tolerances.

Reference [11], offers a better polarization extinction ratio and a better insertion loss than the work presented here. Nevertheless, this device uses thermal tuning and has a difficult calibration that requires many iterations [13].

On the other hand, the work presented in [13] offers a similar polarization extinction ratio as our work and it offers an easy adjustment in just two steps similar as the work proposed here. Their structure will also exhibit the quantum confined Stark effect due to the quantum wells within the structure which is strongly wavelength dependent making adjustment across the C band difficult. Our first devices also show a dependence on wavelength. We do not yet have a good understanding of this but cannot rule out problems with the first round of fabrication. The wavelength dependence might arise due to the Franz-Keldysh (FKE), but a calculation shown in Fig. 7 shows that the Pockels effect is roughly an order of magnitude greater than the FKE effect under the field applied in our structure. We modeled the FKE and the Pockels effect in the structure proposed. The model used to predict the change in the refractive index of the material due to the FKE $\Delta n_{FKE}$ upon the application of an electric field E [V/cm] is described in reference [18] with materials parameters that were taken from references [18–21]. We also modeled the Pockels effect in InGaAsP as described in reference [15].

To compare the phase shift produced by the FKE with the Pockels effect we calculated the phase shift in the phase shifter of the 1x2 MZI. The calculations are summarized in Fig. 7.

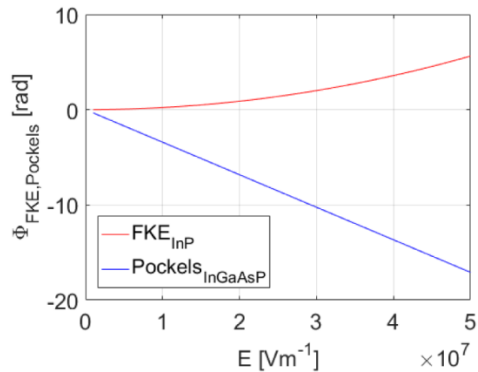

Fig. 7. Phase shift produced in a phase shifter of length L = 2.5 mm upon application of an electric field E. $\Phi_{FKE}$ is the phase due to the FKE (red line) and $\Phi_{Pockels}$ is the phase shift produced by the Pockels effect.

We can see that for a field of about $1 \times 10^7$ Vm$^{-1}$, the Pockels effect is much stronger than the FKE. We will need to fabricate a new set of devices to understand and to control this wavelength dependence.

## 5. Conclusion

We experimentally demonstrated a proof-of-concept of a fabrication tolerant PBS based on a structure that can exhibit both the Pockels effect and the plasma dispersion effect. The device is easily adjustable by controlling two different voltages to produce the required TE/TM splitting. This adjustment can be also used to overcome fabrication tolerances once fabricated. The experimental results show that it can have a polarization extinction ratio better than 15 dB and an on-chip loss around 3.5 dB across the C-band (1530-1565 nm).

## Funding

All calculations were performed on the Lonsdale cluster maintained by the Trinity Centre for High Performance Computing. This work was funded by Science Foundation Ireland through grants 13/TIDA/I2731 and 10/CE/I1853.

## Acknowledgments

We would like to acknowledge Dr. Diarmuid Byrne, and Dr. Michael Gleeson from EBLANA Photonics, Ltd. for fruitful discussions. We would also like to acknowledge Dr. Javier Sánchez Fandiño from Universitat Politècnica de València.